\documentclass[preprint,12pt]{elsarticle}

\usepackage{amssymb}

\journal{}

\begin{document}

\begin{frontmatter}



\title{First-principles study of four quaternary Heusler alloys ZrMnVZ and ZrCoFeZ (Z=Si, Ge) }


\author[mymainaddress]{Huan-Huan Xie}
\author[mymainaddress]{Qiang Gao}
\author[mymainaddress]{Lei Li}
\author[mymainaddress]{Gang Lei}

\author[mymainaddress]{Ge-Yong Mao}

\author[mymainaddress]{Jian-Bo Deng\corref{mycorrespondingauthor}}
\cortext[mycorrespondingauthor]{Corresponding author}
\ead{@lzu.edu.cn}

\address[mymainaddress]{School of Physical Science and Technology, Lanzhou University,
 Lanzhou 730000, People's Republic of China}

\begin{abstract}
We investigate the electronic structure and magnetic properties of four quaternary Heusler alloys ZrMnVZ and ZrCoFeZ (Z=Si, Ge) by using first-principle calculations. It is shown that ZrMnVSi, ZrMnVGe and ZrCoFeSi are half-metallic ferromagnets with considerable half-metallic gaps of 0.14, 0.18 and 0.22 eV, respectively. ZrCoFeGe is a nearly half-metallic, the changes of properties for this alloy under pressure is investigated, the spin polarization of this alloy is 98.99\% at equilibrium lattice constant. 

\end{abstract}


\end{frontmatter}


\section{Introduction}
\label{}
Heusler alloys consist of a large family of intermetallic compounds which attract considerable attention due to the variety of magnetic phenomenon which they present. \cite{1} Half-metallic (HM) magnets are seen as the most promising candidates of high-spin-polarization materials, because their band structure is metallic in one of the two spin channels and semiconducting or insulating in the other one, which results in complete (100\%) spin polarization of electrons at the Fermi level. A number of new half-metallic materials, such as CrAs, NiMnSb, Co$_2$MnAl, etc. \cite{2-1,2-2}, have been initially predicted theoretically by first-principles calculations and later verified by experiments. Among these materials,  the ones with Heusler structure have widely concerned,
 because they can synthesized easily and have high Curie temperature \cite{3-2,3-2,3-3}.

Recently, half-metallic ferromagnetism (HMF)  has been found in Co$_2$ZrSn, Ni$_2$ZrSn \cite{4-1} ,Ni$_2$ZrAl \cite{4-2}, ZrNiSn \cite{4-3}, Ti$_{1-x}$Zr$_x$NiSn \cite{4-4}, ZrCoSb \cite{4-5}. HMF meet all the requirements of spintronics, as a result of their exceptional electronic structure. These materials behave like metals with respect to the electrons of one spin direction and like semiconductors with respect to the electrons of the other spin direction. Zr-based quaternary Heusler ferromagents, ZrFeTiAl, ZrFeTiSi, ZrFeTiGe and ZrNiTiAl,with large HM gaps have been reported lately \cite{5}.

Usually, Heusler alloys have the structural formulas of X$_2$YZ with L2$_1$ structure and XYZ with C1$_1$ structure, where X and Y are transition metals and Z is a main-group element. Many X$_2$YZ and XYZ Heusler alloys have been found to be HM ferromagnets or ferrimagnets \cite{6}. Generally ,X$_1$X$_2$YZ quaternary Heusler compounds crystallize in the LiMgPdSn-type crystal structure \cite{7-1,7-2}. The resulting structure has F4-3m (No. 216) symmetry with Wyckoff positions X$_1$: 4a (0 0 0), X$_2$:4d ($\frac{1}{2}$ $\frac{1}{2}$ $\frac{1}{2}$), Y: 4c ($\frac{1}{4}$ $\frac{1}{4}$ $\frac{1}{4}$), Z: 4b ($\frac{3}{4}$ $\frac{3}{4}$ $\frac{3}{4}$).

In this paper, we explored the electronic structure and magnetic properties of four quaternary Heusler alloys by means of band structure calculation. The results prove that ZrMnVSi, ZrMnVGe and ZrCoFeSi are half-metal, while ZrCoFeGe is a nearly half-metal.

\section{Method of calculations}
\label{}
We have carried out density functional calculation using the scalar relativistic version of the full-potential local-orbital (FPLO) minimum-basis band-structure method \cite{8-1,8-2}. For the present calculations, the site-centered potentials and densities were expanded in spherical harmonic contributions up to $l$$_{max}$=12. The Perdew-Burke-Ernzerhof 96 of the generalized gradient approximation (GGA) was used for exchange-correlation (XC) potential \cite{8-3}.For the irreducible  brillouin zone , we use the $k$ meshes of 20$\times$20$\times$20 for all the calculations. The convergence criteria of self-consistent iterations is set to $10^{-6}$ to the density and $10^{-8}$ Hartree to the total energy per formula unit.

\section{Results and discussions}
\label{}
Quaternary Heusler alloys X$_1$X$_2$YZ have three possible structures \cite{9-1,9-2}. We calculate both the spin-polarized and the nonspin-polarized total energies as a function of volumes for all the ZrMnVZ and ZrCoFeZ in the three possible structures. Table 1 present the optimized lattice constants and the calculated total magnetic moments per formula unit at equilibrium lattices with the most stable structure for ZrMnVZ and ZrCoFeZ. It is shown that ZrMnVSi, ZrMnVGe and ZrCoFeSi are half-metallic ferromagnets, while ZrCoFeGe is nearly half-metallic ferromagnets. 

The calculated total magnetic moment is 1.00 $\mu_B$ per formula unit for ZrCoFeSi (Table 1). There are 25 valence electrons in ZrCoFeSi, and the total magnetic moment of 1.00 $\mu_B$ per formula unit complies with the Slater-Pauling behaviour of HM quaternary Heusler alloys, 
\begin{equation}
M_{tot}=(Z_{tot}-24) \mu_B
\end{equation}   
where Z$_{tot}$ and M$_{tot}$ are the number of total valence electrons and the total magnetic moment. As for the other two HM compounds, there are 20 valence electrons in ZrMnVSi and ZrMnVGe, the total magnetic moment of 2 $\mu_B$ per formula unit complies with the Slater-Pauling behaviour of HM quaternary Heusler alloys,
\begin{equation}
M_{tot}=(Z_{tot}-18) \mu_B
\end{equation}
here Z$_{tot}$ and M$_{tot}$ have the same meaning as previous. 

The total magnetic moment contains four parts (Table 2), As for ZrCoFeSi, the magnetic interaction is ferromagnetic between Co and Fe, are antiferromagnetic between Zr and Co and between Zr and Fe. Different to ZrCoFeSi, the magnetic interaction is ferromagnetic between Zr and V, are antiferromagnetic between Zr and Mn and between V and Mn for both ZrMnVSi and ZrMnVGe.

Fig. 1 show the calculated spin-polarized total density of states (DOS) for ZrMnVSi, ZrMnVGe and ZrCoFeSi at their equilibrium lattices constants. From it, one can see this three HM ferromagnets have considerable HM gaps of 0.14, 0.18 and 0.22 eV, respectively. Near the Fermi level, the spin-up DOS shows a metallic property, while the spin-down DOS has a wide gape, so this three quaternary Heusler alloys keeps an ideal 100\% spin-polarization of conduction electrons at Fermi level.

Fig. 2 shows the calculated band structure of ZrCoFeSi at equilibrium lattice constant as a representation of the three HM compounds . Definitely, ZrCoFeSi exhibits a HM characteristic: the spin-up band structure is metallic, and the energy gap is about 0.64 eV in the spin-down band structure. Fig. 3 shows the calculated results of total and partial DOS of ZrCoFeSi. It can be seen that the total DOS are mainly contributed by the 3d states of Co and Fe atoms in the range of -4 to 2 ev. Near the Fermi level, the 3d states of Co and Fe atoms make a main contribution to the total DOS. Near the Fermi level the 4d states of Zr atoms make the most contribution of the Zr atoms while  the 3p states of Si make the most contribution of the Si atoms. We can also see there is a hybridization between Co-3d state and Fe-3d state around the Fermi level. 

Now, we investigate the HM stability for ZrMnVSi, ZrMnVGe and ZrCoFeSi under uniform strains. Because HM materials are usually used in spintronic devices in the form of thin films or multilayers, the lattice constant will have a change when the films or multilayers are grown on appropriate substrates , and correspondingly the half-metallicity may be destroyed. In order to study the effect of uniform strain (i.e. corresponding to hydrostatic pressure), we calculate the band structure and the total magnetic as a function of lattice constant for ZrMnVSi, ZrMnVGe and ZrCoFeSi. It indicate that the half-metallicity can be retained and the total magnetic moments can be kept an integral number of Bohr magneton (Fig. 4) until the lattice constant are contracted to be 5.86, 5.87 and 5.50 \AA $\ $for ZrMnVSi, ZrMnVGe and ZrCoFeSi, the half-metallicities of ZrMnVSi, ZrMnVGe and ZrCoFeSi can be preserved up to 4\%, 4\%, 8\% compression of lattice constant with respect to their equilibrium lattices, respectively. In addition, we also reveal that  ZrMnVSi, ZrMnVGe and ZrCoFeSi are still HM when their lattice constants are expanded appropriately (Fig. 4).

Next, we will have an investigation on the quaternary  Heusler alloy ZrCoFeGe. There are 25 valence electrons in ZrCoFeSi, and the total magnetic moment of 1.00 $\mu_B$ per formula unit complies with the Slater-Pauling behaviour of HM quaternary Heusler alloys, 
\begin{equation}
M_{tot}=(Z_{tot}-24) \mu_B
\end{equation}   
where Z$_{tot}$ and M$_{tot}$ are the number of total valence electrons and the total magnetic moment. As can be seen in \textbf{Figure 5},  the DOS in the spin-down band structure get through the Fermi level, so
this Heusler alloy is not a half-metal, but a nearly half-metal. According to the definition, the spin polarization (P) can be expressed by,  
\begin{equation}
P=\frac{N\uparrow (\epsilon_F)-N\downarrow (\epsilon_F)}{N\uparrow (\epsilon_F)+N\downarrow (\epsilon_F)}
\end{equation}
where $N\uparrow (\epsilon_F)$ and $N\downarrow (\epsilon_F)$ are the DOS of majority-spin and minority-spin at Fermi level, the values of $N\uparrow (\epsilon_F)$ and $N\downarrow (\epsilon_F)$ can be obtained from \textbf{Figure 5}. Through \textbf{Equation 3}, we can calculated out that the spin polarization of ZrCoFeGe is 98.99\% under its equilibrium lattice constant. 

The alloy ZrCoFeGe has such a high spin polarization that just by applying a very low pressure it may transform into a half-metal. So we investigate the effect of pressure on the properties of ZrCoFeGe. The equation of state (EOS) of Murnaghan \cite{10} is used as a function to study the change of lattice constant under pressure. \textbf{Figure 6} shows the lattice constant of ZrCoFeGe changes under pressure, as can be seen from it, the lattice constant decreases while the pressure increases. We can calculate magnetic moment and band gap as a function of lattice constant, so we can investigate the relationships of magnetic moment and band gap with pressure. Figure 7 and 8 show magnetic moment and band gap change under pressure. As can be seen in \textbf{Figure 7}, The magnetic moment keeps 1 $\mu_B$ when pressure in the range from 0 to 25.53 GPa, if pressure increases continuously until 63.97 GPa, the magnetic moment decrease, beyond 63.97 GPa, the magnetic moment becomes 0. From \textbf{Figure 8}, we can get that  ZrCoFeGe will have band gap just by applying a pressure of 0.20 GPa. In the range from 0.20 GPa to 25.53 GPa, ZrCoFeGe keeps having band gaps, upon 25.53 GPa, ZrCoFeGe has no gaps.  On the whole, ZrCoFeGe keeps being a HM when pressure in the range from 0.2 GPa to 25.53 GPa. It is ferromagnetic from 25.53 GPa to 65.97 GPa,  not ferromagnetic beyond 65.97 GPa,.

\section{Conclusions}
\label{}
In conclusion, we have used first-principle to investigate the electronic  structure and magnetic properties of four quaternary Heusler alloys ZrMnVZ and ZrCoFeZ. It is shown that ZrMnVSi, ZrMnVGe and ZrCoFeSi are HM ferromagnets with considerable HM gaps of 0.14, 0.18 and 0.22 eV, the half-metallicities of ZrMnVSi, ZrMnVGe and ZrCoFeSi can be preserved up to 4\%, 4\%, 8\% compression of lattice constant with respect to their equilibrium lattices. ZrCoFeGe is a nearly HM,  the spin polarization of this alloy is 98.99\% under its equilibrium lattice constant, when apply a pressure of 0.20 Gpa it will transform into a HM. These four Heusler alloys may be promising materials for future spintronic applications.





\bibliographystyle{elsarticle-num}
\bibliography{Reference.bib}







\newpage
\begin{table}[!hbp]

\caption{\label{arttype}The calculated equilibrium lattice constants (a) in \AA, total magnetic moments (m) in $\mu_B$, half-metallic gap (E$_g$) in eV, and physical nature for the four quaternary Heusler alloys at equilibrium lattice constants. }
\footnotesize\rm
\begin{tabular*}{\textwidth}{@{}l*{15}{@{\extracolsep{0pt plus12pt}}l}}
\hline
\hline

Compounds&a(\AA)&m($\mu_B$)&E$_g$(eV)&Physical nature&\\

\hline
ZrMnVSi &6.128&2.00&0.14&Half-metal&\\
ZrMnVGe &6.219&2.00&0.18&Half-meta&\\
ZrCoFeSi &5.973&1.00&0.22&Half-meta&\\
ZrCoFeGe &6.056&1.00&0.00&Nearly half-meta&\\
\hline
\hline
\end{tabular*}
\end{table}

\newpage
\begin{table}[!hbp]
\caption{\label{arttype}The partial magnetic moments of the Heusler alloys ZrMnVSi, ZrMnVGe and ZrCoFeSi under the equilibrium lattice constant.}
\begin{tabular*}{\textwidth}{@{}l*{15}{@{\extracolsep{0pt plus12pt}}l}}
\hline
\hline
	X$_1$X$_2$YZ & m$_{X_1}$($\mu_B$) & m$_{X_2}$ ($\mu_B$) & m$_Y$ ($\mu_B$) & m$_Z$ ($\mu_B$) & $m_{tot}$ ($\mu_B$)\\
\hline
	ZrMnVSi & 0.01 & -0.05 & 2.10 & -0.06 & 2.00\\
    ZrMnVGe & 0.09 & -0.28 & 2.31 & -0.12 & 2.00\\
	ZrCoFeSi & -0.17 & 0.56 & 0.65 & -0.04 & 1.00\\
	\hline
	\hline
\end{tabular*}
\end{table}

\newpage
\begin{figure}[htp]
\centering
\includegraphics[scale=0.6]{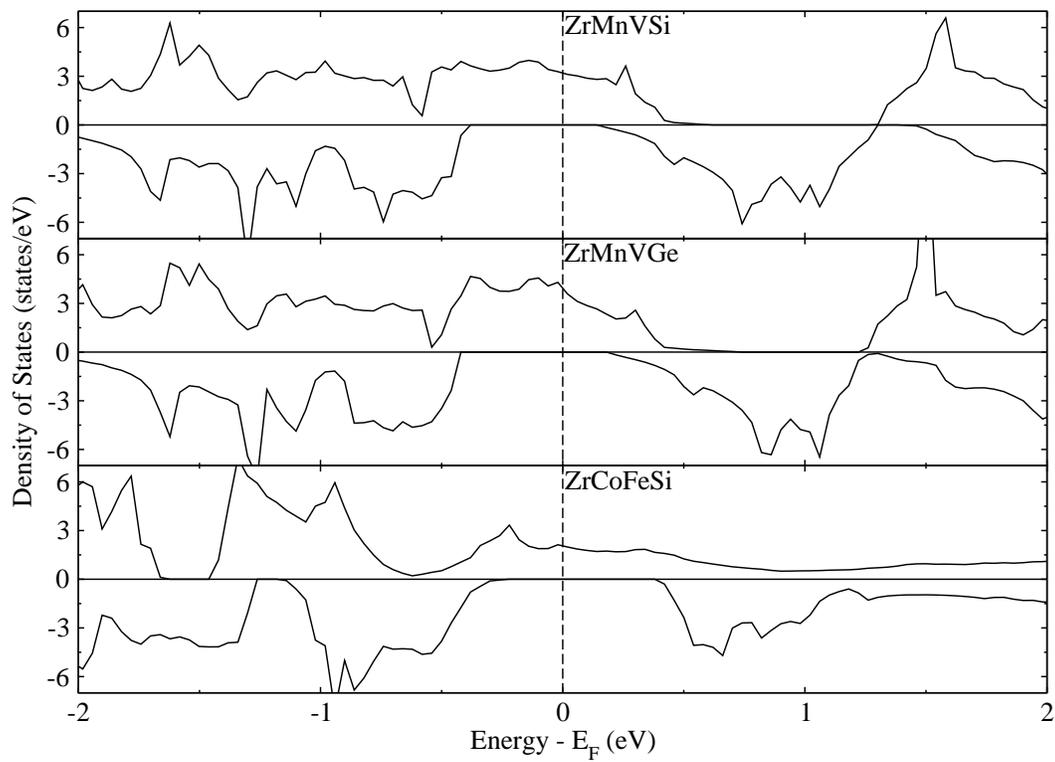}
\caption{The total density of states for ZrMnVSi, ZrMnVGe and ZrCoFeSi at equilibrium lattice constants.}
\label{}
\end{figure}

\newpage
\begin{figure}[htp]
\centering
\includegraphics[scale=0.6,angle=270]{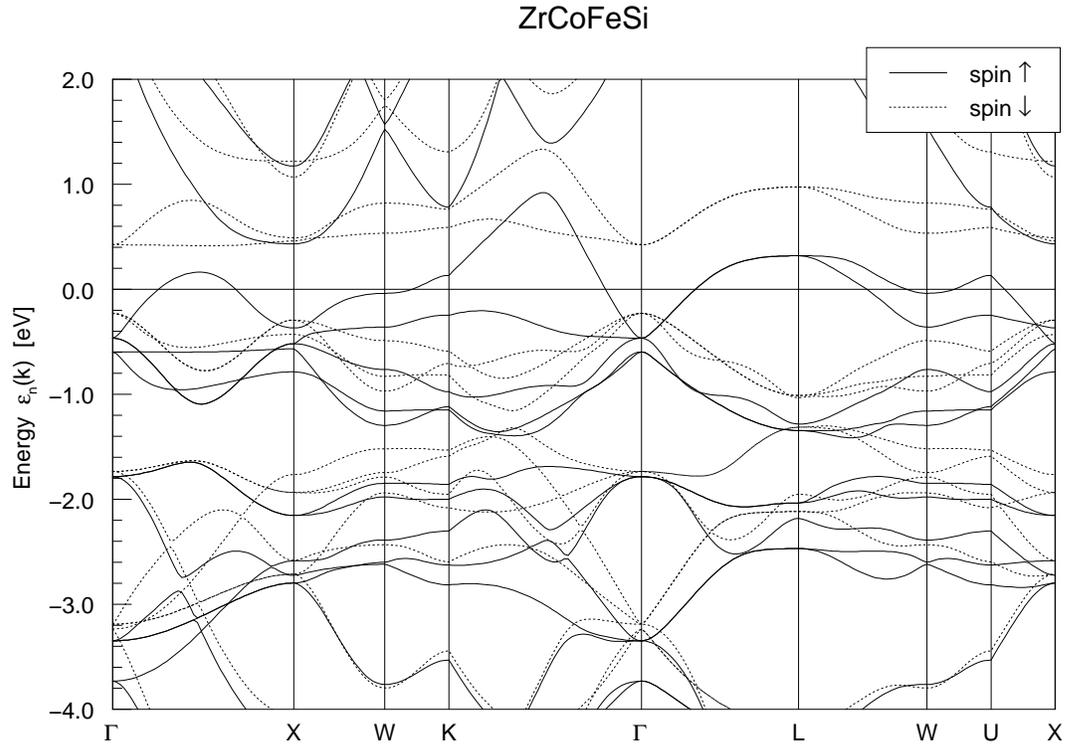}
\caption{Spin-up and spin-down band structure for ZrCoFeSi at the equilibrium lattice constant of 5.973 \AA. Arrows $\uparrow$ and $\downarrow$ represent the spin-up and spin-down states respectively. }
\label{}
\end{figure}

\newpage
\begin{figure}[htp]
\centering
\includegraphics[scale=0.6]{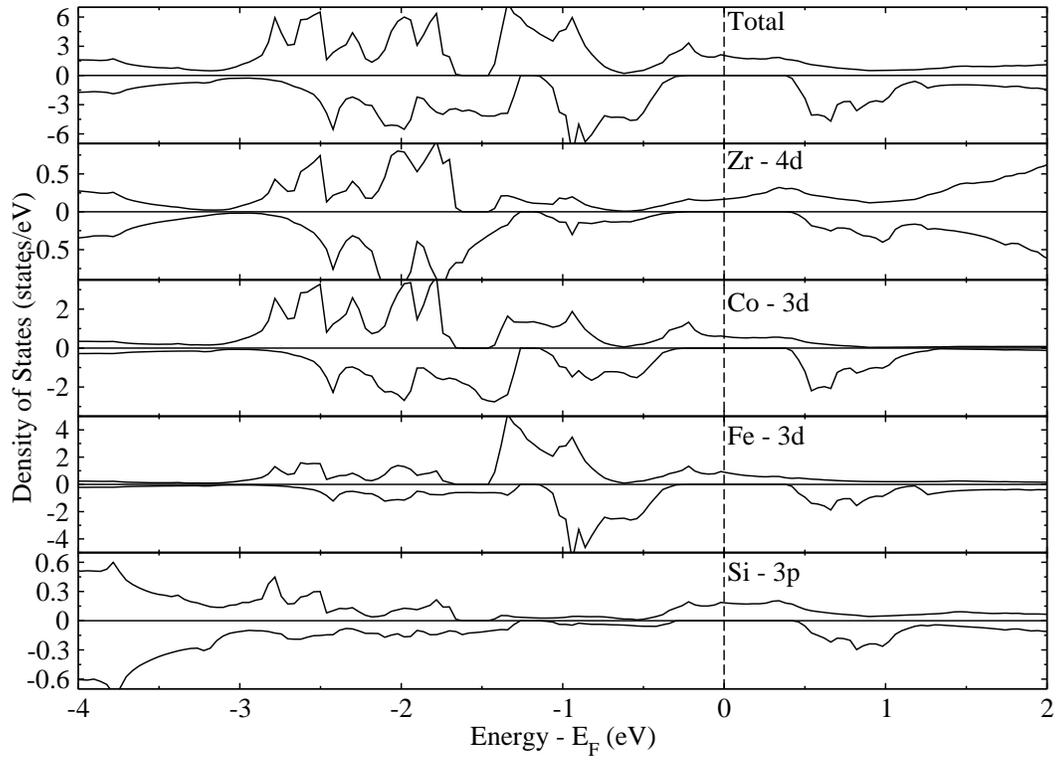}
\caption{The total and partial DOS plots  of ZrCoFeSi at its equilibrium lattice constants. The zero energy value is in correspondence to the Fermi level. Positive values of DOS represent spin-up electrons, negative values represent spin-down electrons. }
\label{}
\end{figure}

\newpage
\begin{figure}[htp]
\centering
\includegraphics[scale=0.6]{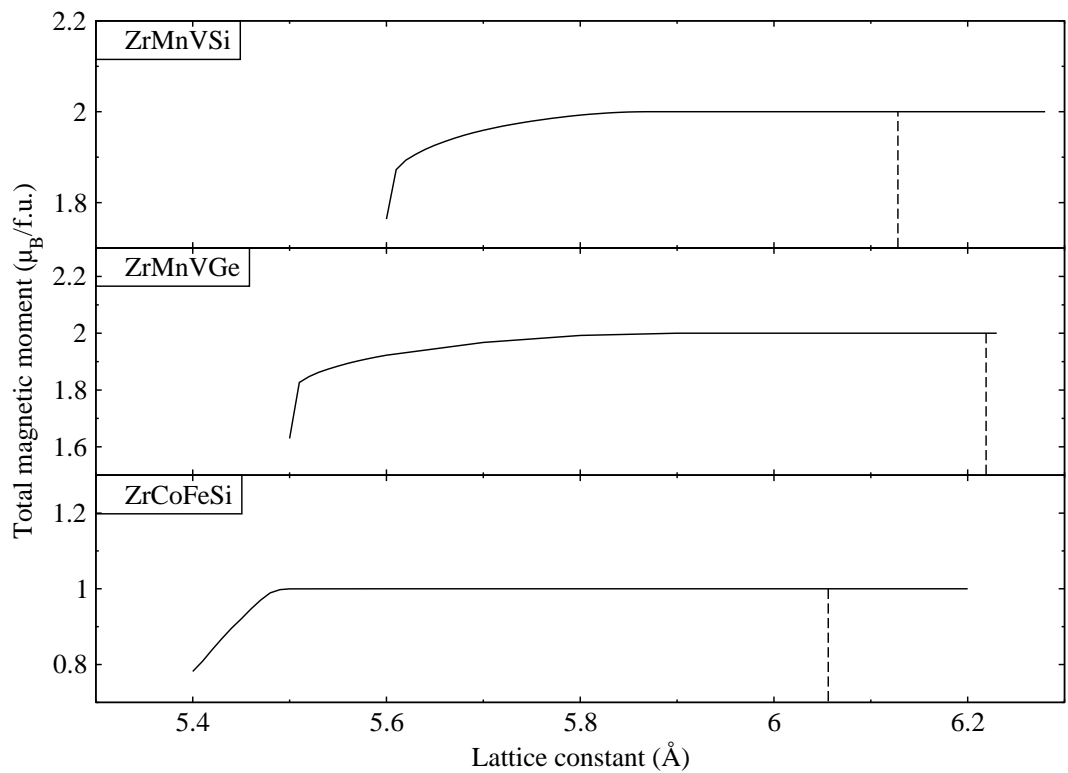}
\caption{Total magnetic moments per formula unit  as a function of lattice constants for ZrMnVSi , ZrMnVGe and ZrCoFeSi. The dot lines indicate the equilibrium lattice constants.}
\label{}
\end{figure}

\newpage
\begin{figure}[htp]
\centering
\includegraphics[scale=0.6]{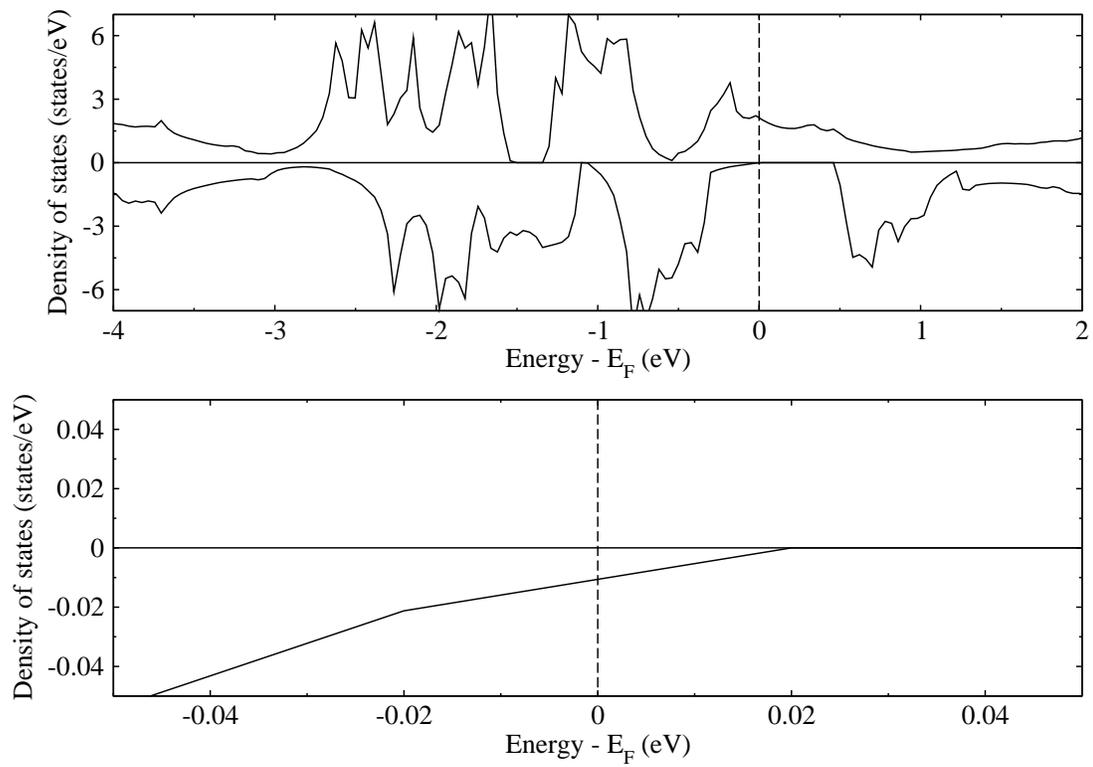}
\caption{The total DOS  and the amplified total DOS near the Fermi level for ZrCoFeGe at equilibrium lattice constant}
\label{}
\end{figure}

\newpage
\begin{figure}[htp]
\centering
\includegraphics[scale=0.6]{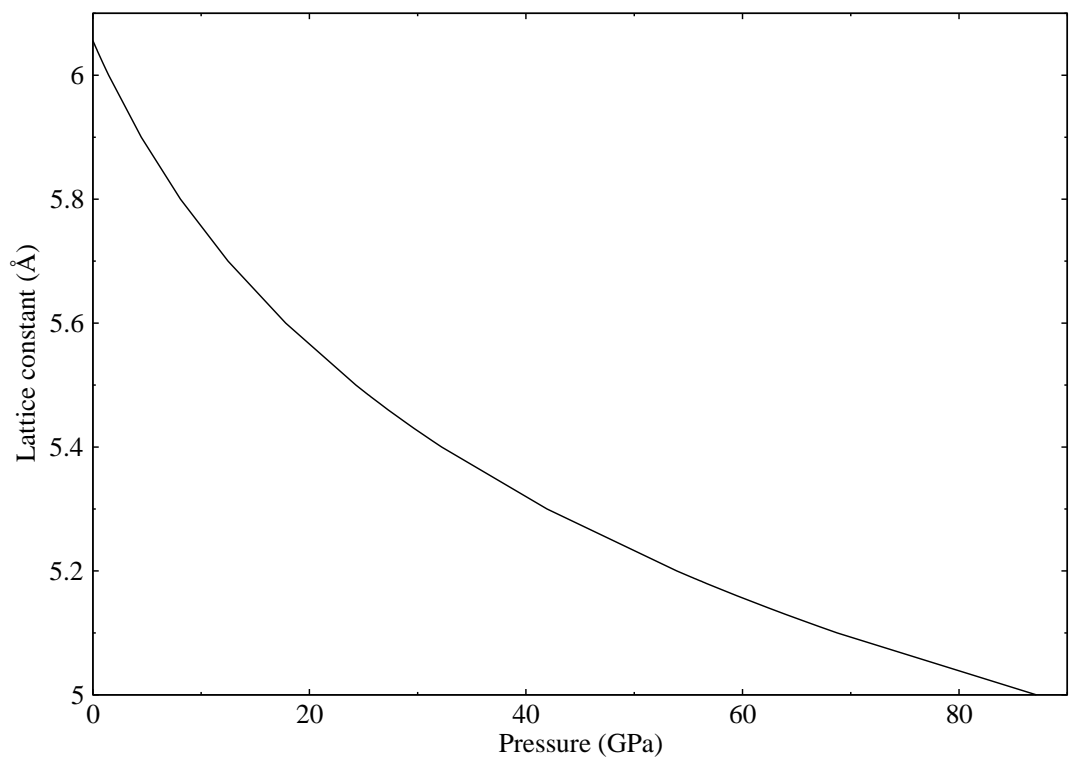}
\caption{The changes of lattice constant for ZrCoFeGe under pressure, zero pressure is in correspondence to equilibrium lattice constant. }
\label{}
\end{figure}

\newpage
\begin{figure}[htp]
\centering
\includegraphics[scale=0.6]{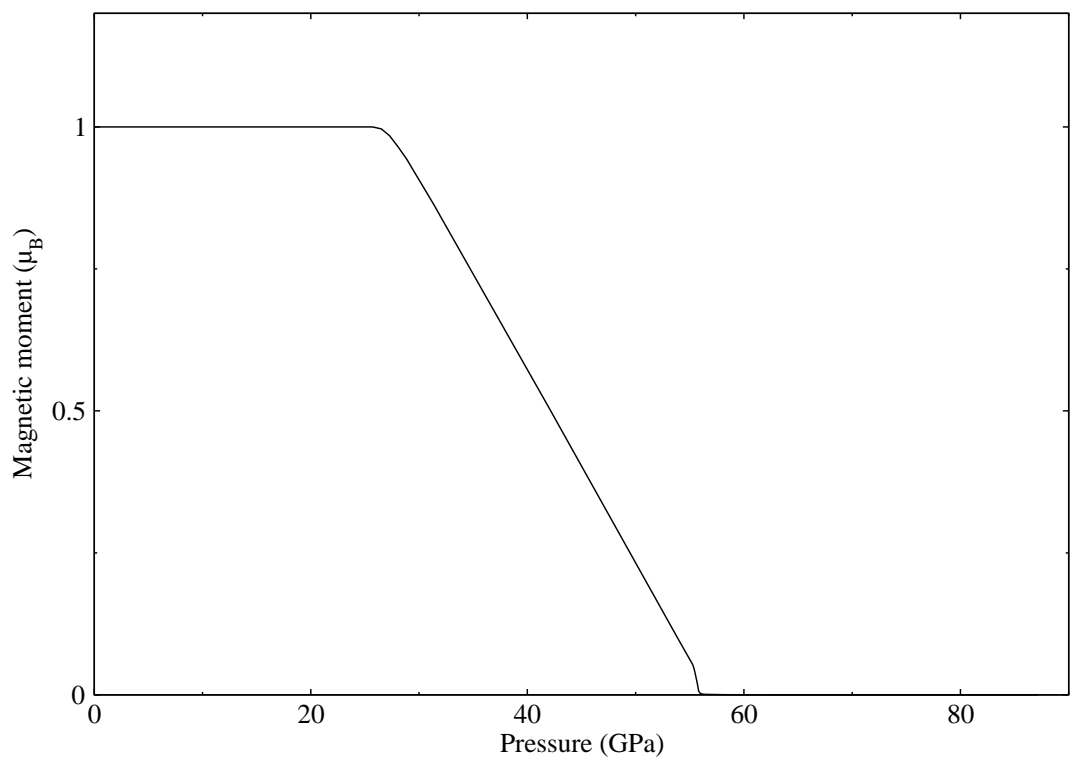}
\caption{The changes of magnetic moment for ZrCoFeGe under pressure. }
\label{}
\end{figure}

\newpage
\begin{figure}[htp]
\centering
\includegraphics[scale=0.6]{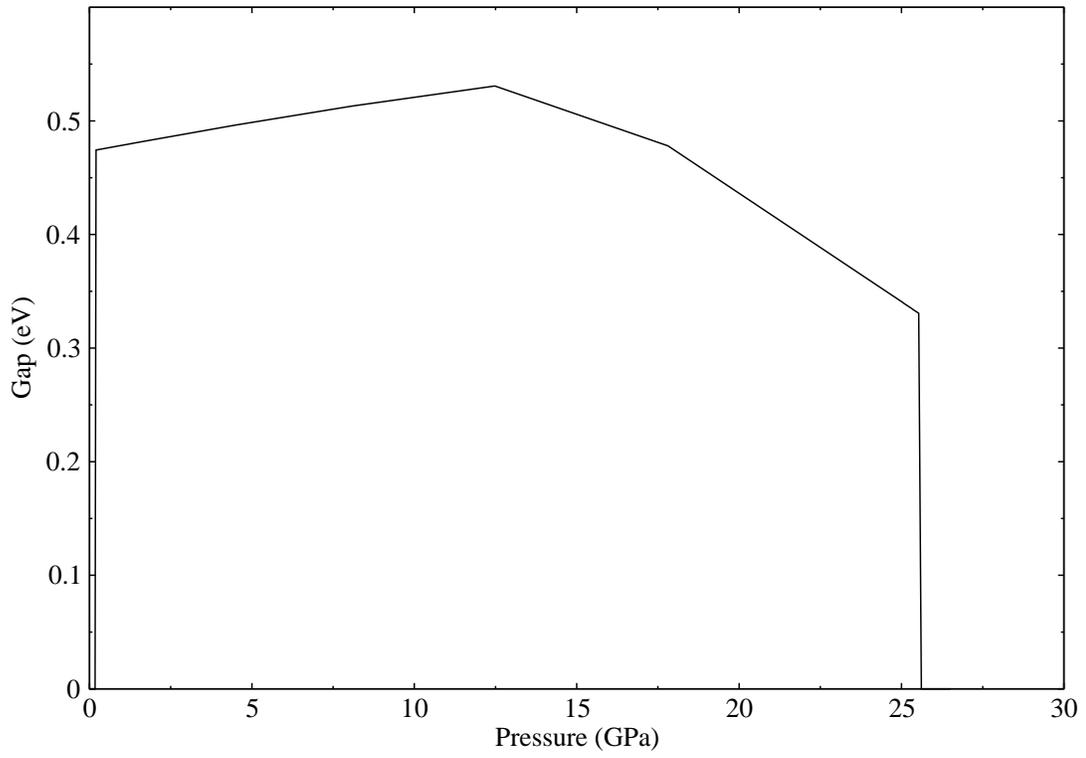}
\caption{The changes of band gap for ZrCoFeGe under pressure.}
\label{}
\end{figure}

\end{document}